\documentclass[10pt]{article}
\usepackage{amsmath}
\usepackage{amssymb}
\usepackage[dvips]{graphicx}
\usepackage[dvips]{epsfig}
\usepackage{color}

 \advance \textwidth by 0.1cm
 \advance \textheight by 0.06cm
\def\##1{{\bf #1}}
\def\=#1{\bar{ #1}}

\def\eps{\epsilon}
\def\epso{\epsilon_0}
\def\muo{\mu_0}
\def\ko{k_0}
\def\lambdao{\lambda_0}
\def\etao{\eta_0}
\def\.{\mbox{ \tiny{$^\bullet$} }}

\def\rll{r_{LL}}
\def\rrr{r_{RR}}

\def\Ezdc{E_{z}^{dc}}

\begin{document}

\begin{center}
{\large {\bf Electrically controlled Bragg resonances of an
ambichiral \\ electro--optic structure: oblique incidence}} \vskip
0.2cm

Mukul Dixit and Akhlesh Lakhtakia

{\small \emph{CATMAS~---~Computational \& Theoretical Materials Sciences Group, \\
Department of Engineering Science \& Mechanics,\\
Pennsylvania State University, University Park, PA 16802--6812, USA.\\
Tel: +1 814 863 4319; Fax: +1 814 865 9974; E--mail: mwd5002@psu.edu, akhlesh@psu.edu}}
\end{center}
\vskip 1em
\hrule
\vskip -0.5em
\noindent{\small The Pockels effect can increase the
effective birefringence of ambichiral, electro--optic rejection filters made of materials with a $\bar{4}2m$ point group symmetry, when a dc electric field is applied parallel to the axis of
nonhomogeneity.  The reflectances and the transmittances of such an
ambichiral structure for obliquely incident plane waves is solvable through a 
boundary--value problem
that is formulated using the frequency--domain Maxwell equations, the
constitutive equations that contain the Pockels effect, and standard
algebraic techniques for handling 4$\times$4 matrix ordinary
differential equations.  The Bragg resonance peaks, for different
circular--polarized--incidence conditions, blueshift as the angle of
incidence increases.  These blueshifts are unaffected by the sign of the dc electric field.  
 \copyright\,
Anita Publications. All rights reserved.
}
\vskip 0.25em
\hrule
\vskip 1em

\noindent{\bf 1 Introduction}

In 1869, Reusch [1] demonstrated that a stack of uniaxial crystalline layers, each rotated about the thickness direction with respect to the
distinguished axis in the layer below by a fixed angle $\Delta\xi$ that is an
integer submultiple of $180^{\circ}$, would transmit circularly
polarized light of one handedness while light of the opposite
handedness would be highly reflected, provided that the stack of layers
is thick enough and the wavelength of the incident light lies in the
Bragg regime.  Such a stack of layers, being periodically piecewise nonhomogeneous in 
the thickness direction, is a Bragg filter.

The optical responses of such structures were analyzed sporadically
after Reusch, mainly in the context of cholesteric liquid crystals [2].  In 2004,
a systematic study of these structures was undertaken [3].  They were classified as 
(i) equichiral, (ii) ambichiral, and (iii) finely chiral structures, depending on the incremental 
angle $\Delta\xi = \pi/q$, $q=2,3,...,$ between two successive layers of the structure [3].  
An equichiral structure $(q=2)$ exhibits the same Bragg resonances for normally incident light of both left-- and 
right--circular polarization states, while an ambichiral structure $(q\geq3)$ exhibits
different Bragg resonances for different circular polarization states.  Finally, finely chiral structures
are classified as those structures in which $q$ approaches infinity while the total thickness remains fixed, so that the
stack of layers resembles a continuous structure like a cholesteric liquid crystal [2] or a chiral sculptured thin film [4].

In 2006, Lakhtakia theoretically analyzed the incorporation of the Pockels
effect in ambichiral structures for small $q\geq3$ [5].  He deduced that layers made of a material with $\bar{4}2m$ point group symmetry 
show increased effective birefringence for a normally incident plane wave, 
on the application of a dc electric field across
the structure in the thickness direction.  Hence, he proposed ambichiral,
electro-optic, circular--polarization--rejection filters to exploit this increase
in effective birefringence, thereby leading to thinner filters than without exploiting
the Pockels effect.

Whereas Lakhtakia restricted the analysis to normally incident plane waves, the analysis in this communication is extended to obliquely incident plane waves.  The dc electric field is still applied in the thickness direction, and the electro--optic material chosen for the layers has a $\bar{4}2m$ point group symmetry.  

The plan of this communication is as follows: Section $2$ contains a description of the ambichiral structure and the optical relative permittivity matrix of an electro--optic material which has $\bar{4}2m$ point group symmetry, along with the formulation of the boundary--value problem to examine the electro--optic responses of the chosen ambichiral structures.  Numerical results  and their interpretive discussion are provided in Section $3$.  Section $4$ contains a brief overview of the key findings.

Throughout this communication, vectors are denoted in boldface; the cartesian unit vectors are represented by $\#{u}_x$, $\#{u}_y$, and $\#{u}_z$; symbols for column vectors and matrixes are decorated by an overbar; and an $\exp(-i\omega t)$ time--dependance is implicit with $i=\sqrt{-1}$, $\omega$ as the angular frequency, and $t$ as time.\\

\noindent{\bf 2 Boundary--Value Problem}

Theoretical analysis of the optical responses of an electro--optic ambichiral structure to an obliquely incident plane wave requires the solution of a boundary--value problem.  The ambichiral structure has $N>>1$ identical layers.
Each layer has a thickness $D$ and extends
infinitely in the transverse (i.e. $xy$) plane.  Hence the total thickness
of the ambichiral structure is $L=ND$ and it occupies the region $0\leq z\leq L$. The halfspaces
$z\leq0$ and $z\geq L$ are assumed to be devoid of any material.

\noindent\emph{2.1 Incident, Reflected and Transmitted Fields}

An arbitrarily polarized plane wave is incident on the
ambichiral structure from the $z\leq0$ halfspace. The
angle of incidence $\theta \in[0,\pi/2)$ of this plane wave on the ambichiral structure with respect to the $z$ axis 
is kept arbitrary.  Consequently, a
reflected plane wave exists in the $z\leq0$ halfspace, and a
transmitted plane wave in the $z\geq L$ halfspace. 

The electric and magnetic field phasors associated with the incident plane wave are
\begin{equation}
\#E_{inc}(\#r) = \#{e}_{inc}(z) \exp{[i \kappa (x
\cos{\psi}+y\sin{\psi})]}\, , \quad z \leq 0\, ,
\end{equation}
\begin{equation}
\#{H}_{inc}(\#r) = \#{h}_{inc}(z) \exp{[i \kappa (x
\cos{\psi}+y\sin{\psi})]}\, , \quad z \leq 0\, ,
\end{equation}
where $\kappa=\ko \sin\theta$; $\ko=2\pi/\lambdao$ is the wavenumber in free space; $\lambdao$ is the free--space wavelength; and $\psi\in[0,2\pi)$ is the direction of propagation of the incident plane wave with respect to
 the $x$ axis in the $xy$ plane.  The plane wave is represented in
 terms of circular--polarization states as
\begin{equation}
\#{e}_{inc}(z)=\left(a_L \frac{i
\#{s}-\#{p}_+}{\sqrt{2}}-a_R\frac{i\#{s}+\#{p}_+}{\sqrt{2}}\right)\exp({i
k_0 z \cos{\theta}}),
\end{equation}
\begin{equation}
\#{h}_{inc}(z)=-i\eta_0^{-1}\left(a_L \frac{i
\#{s}-\#{p}_+}{\sqrt{2}}+a_R\frac{i\#{s}+\#{p}_+}{\sqrt{2}}\right)\exp({i
k_0 z \cos{\theta}}),
\end{equation}
where $\etao=\sqrt{\muo/\epso}$ is the intrinsic impedance of free space, and the quantities $a_L$ and $a_R$ 
are the known amplitudes of the left-- and right--circularly polarized components of the incident plane wave.  The
vectors
\begin{equation}
\#{s}=-\#{u}_x\sin{\psi}+\#{u}_y\cos{\psi},
\end{equation}
\begin{equation}
\#{p}_{\pm}=\mp(\#{u}_x\cos{\psi}+\#{u}_y\sin{\psi})\cos{\theta}+\#{u}_z\sin{\theta},
\end{equation}
are unity in magnitude and are used for notational simplicity.  

Similarly,
the electric and magnetic field phasors associated with the
reflected and transmitted plane waves are
\begin{equation}
\#{E}_{ref}(\#r)  = \#{e}_{ref}(z) \exp{[i \kappa (x
\cos{\psi}+y\sin{\psi})]}\, , \quad z \leq 0\, ,
\end{equation}
\begin{equation}
\#{H}_{ref}(\#r)  = \#{h}_{ref}(z) \exp{[i \kappa (x
\cos{\psi}+y\sin{\psi})]}\, , \quad z \leq 0\, ,
\end{equation}
\begin{equation}
\#{E}_{tr}(\#r)  = \#{e}_{tr}(z) \exp{[i \kappa (x
\cos{\psi}+y\sin{\psi})]}\, , \quad z \geq L\, ,
\end{equation}
\begin{equation}
\#{H}_{tr}(\#r)  = \#{h}_{tr}(z) \exp{[i \kappa (x
\cos{\psi}+y\sin{\psi})]}\, , \quad z \geq L\, ,
\end{equation}
where
\begin{equation}
\#{e}_{ref}(z)=\left(-r_L \frac{i \#{s}-\#{p}_-}{\sqrt{2}}+r_R
\frac{i\#{s}+\#{p}_-}{\sqrt{2}}\right)\exp{(-i k_0 z \cos{\theta})},
\end{equation}
\begin{equation}
\#{h}_{ref}(z)=i\eta_0^{-1}\left(r_L \frac{i
\#{s}-\#{p}_-}{\sqrt{2}}+r_R\frac{i\#{s}+\#{p}_-}{\sqrt{2}}\right)\exp({-i
k_0 z \cos{\theta}}),
\end{equation}
\begin{equation}
\#{e}_{tr}(z)=\left(t_L \frac{i \#{s}-\#{p}_+}{\sqrt{2}}-t_R
\frac{i\#{s}+\#{p}_+}{\sqrt{2}}\right)\exp{\left[i k_0 (z-L) \cos{\theta}\right]},
\end{equation}
\begin{equation}
\#{h}_{tr}(z)=-i\eta_0^{-1}\left(t_L \frac{i
\#{s}-\#{p}_+}{\sqrt{2}}+t_R\frac{i\#{s}+\#{p}_+}{\sqrt{2}}\right)\exp\left[{i
k_0 (z-L) \cos{\theta}}\right].
\end{equation}
The quantities $r_L$ and $r_R$ are the unknown amplitudes of the
reflected planewave components, while $t_L$ and $t_R$ are the unknown
amplitudes of the transmitted planewave components.  

The reflection and
transmission coefficients ($\rll, \rrr$, and so on) can be conveniently defined using
the following $2\times2$ matrix relationships:
\begin{equation}
\left(
        \begin{array}{c}
          r_L \\
          r_R \\
        \end{array}
      \right)=
\left(
  \begin{array}{cc}
    r_{LL} & r_{LR} \\
    r_{RL} & r_{RR} \\
  \end{array}
\right)
\left(
  \begin{array}{c}
    a_L \\
    a_R \\
  \end{array}
\right),
\end{equation}
\begin{equation}
\left(
        \begin{array}{c}
          t_L \\
          t_R \\
        \end{array}
      \right)=
\left(
  \begin{array}{cc}
    t_{LL} & t_{LR} \\
    t_{RL} & t_{RR} \\
  \end{array}
\right) \left(
  \begin{array}{c}
    a_L \\
    a_R \\
  \end{array}
\right).
\end{equation}
Co--polarized coefficients have both subscripts identical, but cross--polarized
coefficients do not.  The square of the magnitude of a reflection or 
transmission coefficient corresponds to
the respective reflectance or transmittance, i.e., $R_{LL}=\big \vert \rll \big \vert^2$ and so on.  Also, constraints are imposed by the principle of conservation of energy as
\begin{equation}
R_{LL}+R_{RL}+T_{LL}+T_{RL} \leq 1,
\end{equation}
\begin{equation}
R_{RR}+R_{LR}+T_{RR}+T_{LR} \leq 1,
\end{equation}
the inequalities turning into equalities when there is no dissipation
of energy inside the ambichiral structure.

\noindent\emph{2.2 Optical Permittivity of the Electro--optic Ambichiral Structure}

An important property desirable for
an ambichiral structure is that it should be transparent over a
certain range of wavelengths, in the present case being the visible and near--infrared regimes.  Therefore, the electro--optic material was considered to be non-dissipative.  Furthermore, the material chosen for the layers has $\bar4 2m$ point group symmetry, examples of
relevant materials being ammonium dihydrogen phosphate and potassium
dihydrogen phosphate, both transparent in the visible and near--infrared regimes [6].

A uniform dc electric field
$\#{E}^{dc}=\Ezdc$ (where $E_z^{dc}$ can be varied in sign and magnitude) is supposed to be applied
across the ambichiral structure by using transparent
indium--tin--oxide electrodes [7].  This electric field is aligned parallel to the thickness direction of the thin
film (the $z$ axis).

The $n^{th}$ layer in the ambichiral structure occupies the region $z_{n-1} \leq z \leq z_n$, $n\in[1,N]$, where $z_m=mD$.  For sufficient generality, the optical relative permittivity matrix of the $n^{th}$ layer is given by
\begin{eqnarray}
\nonumber \bar{\eps}_{r} (z)&=&
\bar{S}_{z}\left(h\xi_n\right)\cdot\bar{R}_{y}(\chi)\cdot\left(
\begin{array}{ccc}
\eps_1 & -r_{63}\,\eps_1^2\,E_{z}^{dc}\,\sin\chi & 0 \\[5pt]
-r_{63}\,\eps_1^2\,E_{z}^{dc}\,\sin\chi & \eps_1 & -r_{41}\,\eps_1\,\eps_3\,E_z^{dc}\,\cos\chi \\[5pt]
0 & -r_{41}\,\eps_1\,\eps_3\,E_{z}^{dc}\,\cos\chi & \eps_3 \\[5pt]
\end{array}
\right)\
\\
& &\cdot\bar{R}_{y}(\chi)\cdot\bar{S}_{z}\left(h\xi_n\right),\quad z_{n-1} \leq z \leq z_n\, , \\[5pt]\nonumber
\end{eqnarray}where $r_{41}$ and $r_{63}$ are the
electro--optic coefficients relevant to the $\bar4 2m$ point group
symmetry; and $\eps_1$ and $\eps_3$ are, respectively, the squares of the ordinary and the extraordinary refractive indexes in the absence of the Pockels effect.  The tilt matrix is defined as
\begin{eqnarray}
\bar{R}_{y}(\chi) = \left(
  \begin{array}{ccc}
    -\sin\chi & 0 & \cos\chi \\
    0 & -1 & 0 \\
    \cos\chi & 0 & \sin\chi \\
  \end{array}
\right), \quad\chi\in[0,\pi/2].
\end{eqnarray}
The rotation matrix
\begin{eqnarray}
\bar{S}_{z}(\zeta)=
 \left(
  \begin{array}{ccc}
    \cos{\zeta} & -\sin{\zeta} & 0 \\
    \sin{\zeta} & \cos{\zeta} & 0 \\
    0 & 0 & 1 \\
  \end{array}
\right),
\end{eqnarray}
indicates rotation about the $z$ axis by an angle of
$\zeta$ with respect to the first layer in the structure.  The
quantity $\xi_n=(n-1)\Delta\xi=(n-1)\pi/q$ with the ratio $N/q$
being an integer. Finally, the parameter $h=+1$ denotes structural right--handedness,
and $h=-1$ is to be used for structural left--handedness.  Equation $(19)$ is correct to the first order in $E_z^{dc}$.  Note that Lakhtakia [5] had treated only the case when $\chi=0$ and $\kappa=0$.

\noindent\emph{2.3 Matrix Ordinary Differential Equation}

The source-free Maxwell
curl postulates
\begin{equation}
\#{\nabla}\times\#{E}(\#{r})=i\omega\muo\#{H}(\#{r}), \quad 0<z<L,
\end{equation}
\begin{equation}
\#{\nabla}\times\#{H}(\#{r})=-i\omega\epso\bar{\epsilon}_r\cdot\#{E}(\#{r}), \quad 0<z<L,
\end{equation}
must hold in the ambichiral structure.  In accordance with the incident plane wave, the Fourier representations
\begin{equation}
\#{E}(\#{r})=[e_x(z)\#{u}_x+e_y(z)\#{u}_y+e_z(z)\#{u}_z]\exp[i\kappa
(x\cos{\psi} + y\sin{\psi})],\quad z\in[0,L],
\end{equation}
\begin{equation}
\#{H}(\#{r})=[h_x(z)\#{u}_x+h_y(z)\#{u}_y+h_z(z)\#{u}_z] \exp[i\kappa
(x\cos{\psi} + y\sin{\psi})],\quad z\in[0,L],
\end{equation}
must be used.

On substituting $(19)$, $(24)$ and $(25)$
in $(22)$ and $(23)$, four ordinary differential equations
and two algebraic equations emerge.  The two algebraic equations are manipulated to eliminate $e_z(z)$ and $h_z(z)$ from the four ordinary
differential equations.  Thereafter, the four ordinary differential
equations are written completely as the $4\times4$ matrix ordinary
differential equation
\begin{eqnarray}
\frac{d}{dz}\bar{f}(z)=i\bar{A}_n\cdot\bar{f}(z), \quad z \in (z_{n-1},z_n),
\end{eqnarray}
where
\begin{eqnarray}
\bar{f}(z)=\left(
         \begin{array}{c}
           e_x(z) \\
           e_y(z) \\
           h_x(z) \\
           h_y(z) \\
         \end{array}
       \right)
\end{eqnarray}
is a column vector and the $4\times4$ matrix
\begin{equation}
\bar{A}_n=\bar{T}_n^{T}\cdot(\bar{A}_{00}+\bar{A}_{01}+\bar{A}_{02}+\bar{A}_{03}+\bar{A}_{s1}+\bar{A}_{s2})\cdot\bar{T}_n.
\end{equation}
Here
\begin{equation}
\bar{T}_n=\left(
                        \begin{array}{cccc}
                      \cos({h\xi_n}) & \sin({h\xi_n}) & 0 & 0 \\
                      -\sin({h\xi_n}) & \cos({h\xi_n}) & 0 & 0 \\
                              0 & 0 & \cos({h\xi_n}) & \sin({h\xi_n}) \\
                                 0 & 0 & -\sin({h\xi_n}) & \cos({h\xi_n}) \\
                              \end{array}
                          \right),
\end{equation}
\begin{equation}
\bar{A}_{00}=\left(
                        \begin{array}{cccc}
                      0 & 0 & 0 & \omega\muo \\
                      0 & 0 & -\omega\muo & 0 \\
                      0 & -\omega\epso\eps_1 & 0 & 0 \\
                      \omega\epso\eps_d & 0 & 0 & 0 \\
                              \end{array}
                          \right),
\end{equation}
\begin{equation}
\bar{A}_{01}=\kappa \alpha_3 \bar{C}_1,
\end{equation}
\begin{equation}
\bar{A}_{02}=\frac{\kappa^2\eps_d}{\omega\epso\eps_1\eps_3}\bar{C}_3,
\end{equation}
\begin{equation}
\bar{A}_{03}=-\frac{\kappa^2}{\omega\muo}\bar{C}_4,
\end{equation}
\begin{equation}
\bar{A}_{s1}=-\omega\epso\left(
                        \begin{array}{cccc}
                      0 & 0 & 0 & 0 \\
                      0 & 0 & 0 & 0 \\
                      \eps_e & 0 & 0 & 0 \\
                      0 & -\eps_e & 0 & 0 \\
                              \end{array}
                          \right),
\end{equation}
\begin{equation}
\bar{A}_{s2}=\frac{\kappa\eps_f}{\eps_3}\bar{C}_2,
\end{equation}
\begin{equation}
\bar{C}_{1}=\left(
                        \begin{array}{cccc}
                      \cos({h\xi_n-\psi}) & 0 & 0 & 0 \\
                      -\sin({h\xi_n-\psi}) & 0 & 0 & 0 \\
                      0 & 0 & 0 & 0 \\
                      0 & 0 & \sin({h\xi_n-\psi}) & \cos({h\xi_n-\psi}) \\
                              \end{array}
                          \right),
\end{equation}
\begin{equation}
\bar{C}_{2}=\left(
                        \begin{array}{cccc}
                      0 & -\cos({h\xi_n-\psi}) & 0 & 0 \\
                      0 & \sin({h\xi_n-\psi}) & 0 & 0 \\
                      0 & 0 & \sin({h\xi_n-\psi}) & \cos({h\xi_n-\psi}) \\
                      0 & 0 & 0 & 0 \\
                              \end{array}
                          \right),
\end{equation}
\begin{equation}
\bar{C}_{3}=\left(
                        \begin{array}{cccc}
                      0 & 0 & -\cos({h\xi_n-\psi})\sin({h\xi_n-\psi}) & -\cos^2{({h\xi_n-\psi})} \\
                      0 & 0 &  \sin^2{({h\xi_n-\psi})}  &  \cos({h\xi_n-\psi})  \sin({h\xi_n-\psi})  \\
                      0 & 0 & 0 & 0 \\
                      0 & 0 & 0 & 0 \\
                              \end{array}
                          \right),
\end{equation}
\begin{equation}
\bar{C}_{4}=\left(
                        \begin{array}{cccc}
                        0 & 0 & 0 & 0 \\
                        0 & 0 & 0 & 0 \\
                        -\cos({h\xi_n-\psi})\sin({h\xi_n-\psi}) & -\cos^2{({h\xi_n-\psi})} & 0 & 0 \\
                        \sin^2{({h\xi_n-\psi})} & \cos({h\xi_n-\psi})  \sin({h\xi_n-\psi}) &  0  & 0  \\
                                  \end{array}
                          \right),
\end{equation}
\begin{equation}
\eps_d=\frac{\eps_1\eps_3}{\eps_1\cos^2{\chi}+\eps_3\sin^2{\chi}},
\end{equation}
\begin{equation}
\eps_e=E_z^{dc}\eps_1\eps_d (r_{41}\cos^2{\chi}-r_{63}\sin^2{\chi}),
\end{equation}
\begin{equation}
\eps_f=E_z^{dc}\eps_d\cos{\chi}\sin{\chi} (r_{41}\eps_3+r_{63}\eps_1),
\end{equation}
\begin{equation}
\alpha_3=\frac{\eps_d\sin({2\chi})(\eps_1-\eps_3)}{2\eps_1\eps_3}.
\end{equation}

The solution of the matrix ordinary differential equation $(26)$ is
\begin{equation}
\bar{f}(z_n)=\exp[i\bar{A}_nD]\cdot\bar{f}(z_{n-1}).
\end{equation}
Hence, the method to obtain the unknown reflection and transmission amplitudes involves the transfer equation
\begin{equation}
\bar{f}(z_N)=\bar{M}\cdot \bar{f}(z_{0}),
\end{equation}
where the transfer matrix 
\begin{equation}
\bar{M}=\exp [i \bar{A}_ND]\cdot\exp [i \bar{A}_{N-1}D]...\exp [i
\bar{A}_2D]\cdot\exp [i \bar{A}_1D]
\end{equation}
relates the tangential field components on the entry and exit surfaces of the ambichiral structure of thickness $L$ because $z_N=L$ and $z_0=0$.

As the tangential components of $\#{E}(\#{r})$ and $\#{H}(\#{r})$ are continuous
across the planes $z=0$ and $z=L$, the boundary values
\begin{equation}
\bar{f}(0)=\frac{\bar{K}(\theta,\psi)}{\sqrt{2}}\cdot\left(\begin{array}{c}
           i(a_L-a_R) \\
           -(a_L+a_R) \\
           -i(r_L-r_R) \\
           r_L+r_R \\
         \end{array}
       \right),
\end{equation}
\begin{equation}
\bar{f}(L)=
\frac{\bar{K}(\theta,\psi)}{\sqrt{2}}\cdot\left(\begin{array}{c}
           i(t_L-t_R) \\
           -(t_L+t_R) \\
           0 \\
           0 \\
         \end{array}
       \right),
\end{equation}
can be used in $(45)$ with the $4\times4$ matrix
\begin{eqnarray}
\bar{K}(\theta,\psi)=\left(\begin{array}{cccc}
	-\sin{\psi} & -\cos{\theta}\cos{\psi} & -\sin{\psi} & \cos{\theta}\cos{\psi}\\
	  \cos{\psi} & -\cos{\theta}\sin{\psi} & \cos{\psi} & \cos{\theta}\sin{\psi}\\
	 -\eta_0^{-1}\cos{\theta}\cos{\psi} & \eta_0^{-1}\sin{\psi} & \eta_0^{-1}\cos{\theta}\cos{\psi} & \eta_0^{-1}\sin{\psi}\\
	  -\eta_0^{-1}\cos{\theta}\sin{\psi} & -\eta_0^{-1}\cos{\psi} & \eta_0^{-1}\cos{\theta}\sin{\psi} & -\eta_0^{-1}\cos{\psi}\\
	  \end{array}\right).
\end{eqnarray}

The boundary--value problem thus gets simplified to four simultaneous, linear algebraic equations
which can be represented in matrix form as
\begin{equation}
\left(\begin{array}{c}
           i(t_L-t_R) \\
           -(t_L+t_R) \\
           0 \\
           0 \\
         \end{array}
       \right)=\bar{K}^{-1}(\theta,\psi)\cdot\bar{M}\cdot\bar{K}(\theta,\psi)\cdot\left(\begin{array}{c}
           i(a_L-a_R) \\
           -(a_L+a_R) \\
           -i(r_L-r_R) \\
           r_L+r_R \\
         \end{array}
       \right).
\end{equation}
These four equations were solved by matrix manipulation to compute the reflection
and transmission coefficients for arbitrary $\theta$, $E_z^{dc}$, and $\chi$.  Care was taken to validate the results against known results for normal--incidence conditions [5], for both electro--optic and non--electro--optic cases.\\

\noindent{\bf 3 Numerical Results and Discussion}

The following parameter was used for the analysis of the optical response of the electro--optic ambichiral structure to an obliquely incident plane wave [3]:
\begin{equation}
G=q D (\eps_1^{1/2} + \eps_d^{1/2})/\lambdao .
\end{equation}
The four reflectances and the four transmittances were calculated for a range of values of $\theta$, $E_z^{dc}$, and $\chi$ as functions of $G$.  The constitutive parameters used are that of ammonium dihydrogen phosphate at
$\lambdao= 546$~nm: $\eps_1 = 1.53^2$, $\eps_3= 1.483^2$, $r_{41} =
24.5 \times 10^{-12}$~m V$^{-1}$, and $r_{63}=8.5 \times 10^{-12}$~m
V$^{-1}$ [8,9].  Also, the following structural parameters were selected: $\Omega = 500$~nm (where $\Omega =qD$), $q=3$, $h=1$, and $N=24q$.

\begin{center}
\begin{figure}[!htb]
\centering \psfull
\epsfig{file=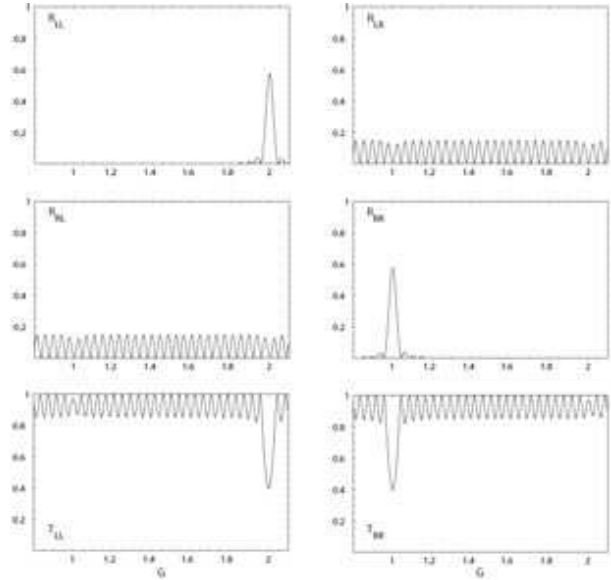, width=8cm}
\caption{
{\small Reflectances and transmittances of a $\bar42m$ point group symmetry ambichiral electro--optic structure plotted as a function of $G$ for a normally incident plane wave.  Bragg resonance peaks occur at $G=1$
for incident right--circularly polarized plane waves and at $G=2$ for left--circularly polarized plane waves.  As $T_{LR}$ and $T_{RL}$ are virtually null--valued, their plots were not included in this figure.  The following parameters were used: $E_z^{dc}=0$, $\theta=0^\circ$, $\chi=0^\circ$.
}}
\end{figure}
\end{center}

\vskip -0.5cm

Figure~1 displays the plots of reflectances and transmittances as  functions of $G$ for a normally incident plane wave when no dc electric field is applied.  Bragg resonance peaks occur at $G=1$
for incident right--circularly polarized plane waves and at $G=2$ for left--circularly polarized plane waves.  Figure~2 displays the reflectances and transmittances of the same structure for $E_z^{dc}=1.5\times 10^9$~V~m$^{-1}$.  Clearly, the intensities of the Bragg resonance peaks increase as the magnitude of $E_z^{dc}>0$ increases.  The same results were obtained for negative $E_z^{dc}$.

\begin{center}
\begin{figure}[!htb]
\centering \psfull
\epsfig{file=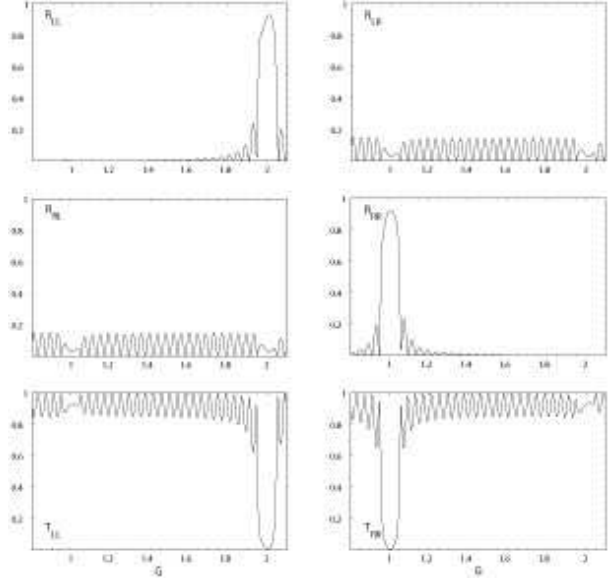, width=8cm}
\caption{
{\small Same as Figure 1, except that $E_z^{dc}=1.5 \times 10^{9}$~V m$^{-1}$.
}}
\end{figure}
\end{center}


\begin{center}
\begin{figure}[!htb]
\centering \psfull
\epsfig{file=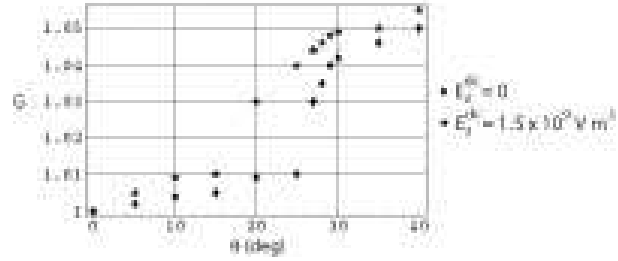, width=8cm}
\caption{
{\small
Plot of $G$ for maximum $R_{RR}$ with respect to $\theta$ (in degree) for $E_z^{dc}=0$ and $E_z^{dc}=1.5 \times 10^{9}$~V m$^{-1}$.
}}
\end{figure}
\end{center}

\vskip -0.5cm

Figure 3 shows plots of the value of $G$ for maximum $R_{RR}$ with respect to $\theta$ in degree for $E_z^{dc}=0$ and $E_z^{dc}=1.5\times 10^9$~V~m$^{-1}$.  The plot for $E_z^{dc}=0$ can be expressed as
\begin{equation}
G_{RR1}(\theta)\simeq1-22.5734\,\theta+13.8206\,\theta^2-3.51473\,\theta^3.
\end{equation}
Similarly, the plot for $E_z^{dc}=1.5\times 10^9$~V~m$^{-1}$ can be expressed as
\begin{equation}
G_{RR2}(\theta)=1+8.07591\,\theta-4.99072\,\theta^2+1.28391\,\theta^3.
\end{equation}
Figure 4 contains plots of the value of $G$ for maximum $R_{LL}$ with respect to $\theta$ in degree for $E_z^{dc}=0$ and $E_z^{dc}=1.5\times 10^9$~V~m$^{-1}$.  The plot for $E_z^{dc}=0$ can be expressed as
\begin{equation}
G_{LL1}(\theta)\simeq2-7.65916\,\theta+4.54041\,\theta^2-1.10564\,\theta^3.
\end{equation}
Similarly, the plot for when $E_z^{dc}=1.5\times 10^9$~V~m$^{-1}$ can be expressed as
\begin{equation}
G_{LL2}(\theta)=2+8.07591\,\theta-4.99072\,\theta^2+1.28391\,\theta^3.
\end{equation}
Such parametric equations can be used to design circular--polarization--rejection filters.

\begin{center}
\begin{figure}[!htb]
\centering \psfull
\epsfig{file=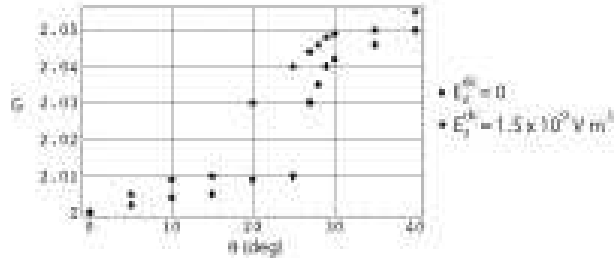, width=8cm}
\caption{
{\small
 Plot of $G$ for maximum $R_{LL}$ with respect to $\theta$ (in degree) for $E_z^{dc}=0$ and $E_z^{dc}=1.5 \times 10^{9}$~V m$^{-1}$.
}}
\end{figure}
\end{center}

\begin{center}
\begin{figure}[!htb]
\centering \psfull
\epsfig{file=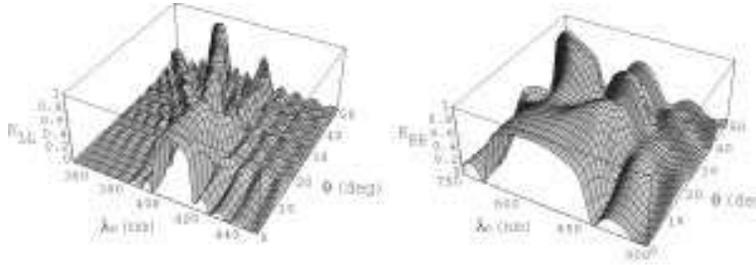, width=10cm}
\caption{
{\small
Plots of $R_{LL}$ and $R_{RR}$ with respect to $\lambdao$ (in nm) and $\theta$ (in degree) for $E_z^{dc}=1.5\times 10^{9}$~V m$^{-1}$ and $\chi=0^\circ$.
}}
\end{figure}
\end{center}

\vskip -0.5cm

Figure 5 shows three-dimensional plots of $R_{LL}$ and $R_{RR}$ with respect to $\lambdao$ and $\theta$, for $E_z^{dc}=1.5 \times 10^{9}$~V m$^{-1}$ and $\chi=0^\circ$.  The Bragg resonance peaks for both incident right-- and left--circularly polarized plane waves shift to the left with increasing $\theta$, i.e. a blueshift in the Bragg resonances can be seen.

The intensities of the Bragg resonance peaks increase as the magnitude of $E_z^{dc}$ increases.  Hence, the application of a more intense dc electric field leads to better Bragg filters for circularly polarized plane waves [5].  Also, the same results were obtained for negative $E_z^{dc}$.  This is further supported by Figure~6, wherein $R_{LL}$ and $R_{RR}$ are plotted with respect to $\lambdao$ (in nm) and $E_z^{dc}$ (in V~m$^{-1}$) for $\theta=15^\circ$ and $\chi=0^\circ$.

Figure 7 contains three-dimensional plots of $R_{LL}$ and $R_{RR}$ with respect to $\lambdao$ and $\chi$, for $E_z^{dc}=\pm1.5 \times 10^{9}$~V m$^{-1}$ and $\theta=15^\circ$.  The Bragg resonance peak for left--circularly polarized plane waves shifts to the right: from $\lambdao=407$ nm at $\chi=0^\circ$ to $\lambdao=414$ nm at $\chi=90^\circ$. Similarly, the Bragg resonance peak for right--circularly polarized plane waves shifts to the right: from $\lambdao=808$ nm at $\chi=0^\circ$ to $\lambdao=826$ nm at $\chi=90^\circ$.  The magnitudes of the Bragg resonance peaks decrease as $\chi$ increases from $0^\circ$.  Around $\chi=50^\circ$ the magnitude of the peaks start to increase again up till $\chi=90^\circ$.  Hence, the lowest possible $\chi$ is most desirable for better Bragg filters if the objective is to achieve broadband performance.  Mid--range values of $\chi$ must be avoided.

\begin{center}
\begin{figure}[!t]
\centering \psfull
\epsfig{file=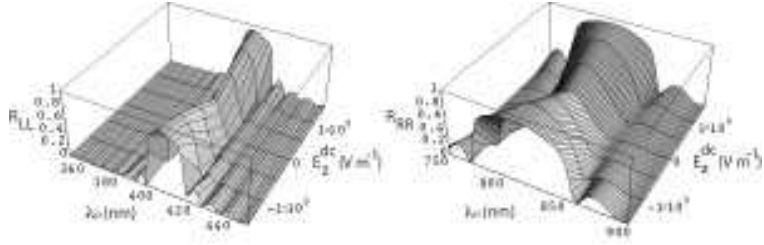, width=10cm}
\caption{
{\small
Plots of $R_{LL}$ and $R_{RR}$ with respect to $\lambdao$ (in nm) and $E_z^{dc}$ (in V m$^{-1}$) for $\theta=15^\circ$ and $\chi=0^\circ$.
}}
\end{figure}
\end{center}

\begin{center}
\begin{figure}[!htb]
\centering \psfull
\epsfig{file=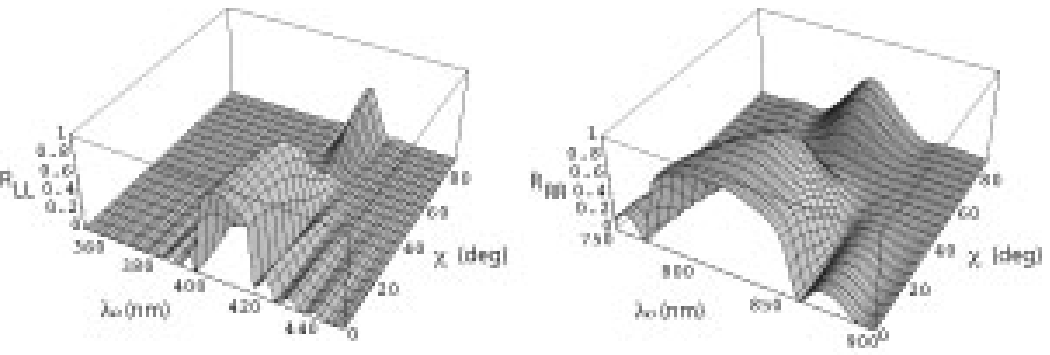, width=10cm}
\caption{
{\small
Plots of $R_{LL}$ and $R_{RR}$ with respect to $\lambdao$ (in nm) and $\chi$ (in degree) for $E_z^{dc}=1.5 \times 10^{9}$~V m$^{-1}$ and $\theta=15^\circ$.
}}
\end{figure}
\end{center}

\vskip-0.5cm
\noindent{\bf 4 Conclusion}

We have shown here that the Pockels effect can be exploit to control the performances
 of ambichiral, electro--optic rejection filters made of materials with a $\bar{4}2m$ point group symmetry, by applying a dc electric field  parallel to the axis of
nonhomogeneity.  The reflectances and the transmittances of such an
ambichiral structure for obliquely incident plane waves were obtained by solving a boundary--value problem that was formulated using the frequency--domain Maxwell equations, the
constitutive equations that contain the Pockels effect, and standard
algebraic techniques for handling 4$\times$4 matrix ordinary
differential equations.   The main results obtained  are as follows:

\begin{enumerate}
\item
The Bragg resonance peaks for both right-- and left--circularly polarized plane waves blueshift as the angle of incidence $\theta$ increases.  The same phenomenon is observed with or without a dc electric field.
\item
The reversal of direction of $E_z^{dc}$ does not affect the blueshift phenomenon in the Bragg resonance peaks, $\theta$ and $\chi$ remaining constant, at least for the chosen point group symmetry.
\item
As the magnitude of $E_z^{dc}$ increases, the intensities of the Bragg peaks deepen, thereby leading to better Bragg filters for circularly polarized states.
\item
The behavior of the peak--reflectance $G$ with varying $\theta$ can be modeled into equations that can be used to design circular--polarization--rejection filters.
\item
The Bragg resonance peaks redshift as $\chi$ increases.  However, as $\chi$ increases, the intensities of the Bragg peaks first decrease to a certain point and then increase.  Hence, Bragg filters with the lowest possible $\chi$ for broadband performance.  Mid--range values of $\chi$ must be avoided while designing Bragg filters.
\end{enumerate}
We conclude by observing that the insertion of a central phase defect in the electro--optic ambichiral
structure would be useful for making electrically tunable narrowband and ultranarrowband filters for circularly polarized plane waves [10--12].

\noindent\emph{\small This paper is dedicated to the affectionate memory
of Prof. Prasad Khastgir, who lit the path of physics for several generations
of students.}\\

\noindent{\bf References}
{\small
\begin{enumerate}

\item
Reusch E, 
\textit{Ann Phys Chem Lpz\/}, 138 (1869) 628.

\item
Collings P J,  \textit{Liquid crystals.\/} (Princeton University Press, Princeton, NJ, USA), 1990.

\item
Hodgkinson I J, Lakhtakia A, Wu Q H, De Silva L, McCall M W,
\textit{Opt Commun\/}, 239 (2004) 353.

\item
Lakhtakia A, Messier R,
\textit{Sculptured thin films: Nanoengineered morphology and optics.\/}
 (SPIE Press, Bellingham, WA, USA), 2005.

\item
Lakhtakia A,
\textit{Phys Lett A\/}, 354 (2006) 330.

\item
http://www.kayelaby.npl.co.uk/general\_physics/2\_5/2\_5\_8.html (18 April 2007).

\item
Osikowicz W, Crispin X, Tengstedt C, Lindell L, Kugler T, Salaneck W R,
\textit{Appl Phys Lett\/}, 85 (2004) 1616.

\item
Boyd R W, \textit{Nonlinear optics.\/}  (Academic Press, San Diego, CA, USA), 1992.

\item
Horn M W, Pickett M D, Messier R, Lakhtakia A, \textit{Nanotechnology\/}, 15 (2004) 303.

\item
Lakhtakia A, \textit{Asian J Phys\/}, 15 (2006) 275.

\item
Lakhtakia A, \textit{J Eur Opt Soc: Rapid Publ\/}, 1 (2006) 06006.

\item
Lakhtakia A, \textit{Opt Commun\/}, 275 (2007) 283.

\end{enumerate}

\end{document}